\documentclass[12pt,aip,jcp,prreprint,noshowkeys,superscriptaddress]{revtex4-1}

\usepackage[fleqn]{amsmath}
\usepackage{bm}
\usepackage{amsmath, amssymb} 
\usepackage{physics}
\usepackage{graphicx}
\usepackage{float}
\usepackage{hyperref}
\usepackage{mathptmx}
\usepackage[version=4]{mhchem}
\usepackage{siunitx}
\usepackage[miktex]{gnuplottex}
\usepackage{subcaption}

\usepackage{gnuplottex}

\usepackage{natbib}
\usepackage{xcolor}

\newcommand{\UCAM}{Yusuf Hamied Department of Chemistry, University of Cambridge, Lensfield Road, Cambridge, CB2 1EW, U.K.}
\newcommand{\UNB}{Department of Chemistry, University of New Brunswick, Fredericton, Canada}
\newcommand{\UNBMATH}{Department of Mathematics \& Statistics, University of New Brunswick, Fredericton, Canada}
\newcommand{\UGHENT}{Department of Chemistry, Ghent University, Ghent, Belgium}
\newcommand{\UCL}{Department of Chemistry, University College London, London, WC1H 0AJ, U.K.}

\newcommand{\textoverline}[1]{$\overline{\mbox{#1}}$}

\hypersetup{
    colorlinks=true,
    linkcolor=blue,
    filecolor=magenta,      
    urlcolor=cyan,
    citecolor=blue,
    }
\graphicspath{./images/}
\begin{document}

\title{Spin-Generator Coordinate Method for Electronic Structure}
\author{Amir~Ayati}
\affiliation{\UNB}
\author{Hugh~G.~A.~Burton}
\affiliation{\UCAM}
\affiliation{\UCL}
\author{Patrick~Bultinck}
\affiliation{\UGHENT}
\author{Stijn~De~Baerdemacker}
\affiliation{\UNB}
\affiliation{\UNBMATH}
\date{\today}

\begin{abstract}
We present a new application of the Generator Coordinate Method (GCM) as an electronic structure method for strong electron correlation in molecular systems.  We identify spin fluctuations as an important generator coordinate responsible for strong static electron correlation that is associated with bond-breaking processes. Spin-constrained Unrestricted HF (c-UHF) states are used to define a manifold of basis states for the Hill--Wheeler equations,  which are discretized and solved as a non-orthogonal configuration interaction (NOCI) expansion. The method was tested on two-electron systems that are dominated by static and/or dynamic correlations. In a minimal basis set for \ce{H2}, the resulting GCM quickly captures the ground-state full configuration interaction energy with just a few c-UHF states, whereas second-order perturbation theory on top of the GCM is needed to recover over $90\%$ of the correlation energy in the cc-pVDZ basis set.
\end{abstract} 
\maketitle

\raggedbottom

\section{Introduction}

Strong electron correlation is one of the outstanding challenges in electronic structure theory.\cite{Hirata2012, Lowdin1995, Lowdin1965} L\"owdin defined electron correlation energy as the energy that Hartree--Fock (HF) theory fails to include with respect to the exact non-relativistic energy\cite{Lowdin1958} and is typically categorized into dynamic and static (or non-dynamic) correlation.\cite{Mok1996, Handy2001, Cremer2001, Becke2013, Crittenden2013, Tsuchimochi2014, Wallace2014, Hollett2016, RamosCordoba2016, BenavidesRiveros2017, ViaNadal2019} Dynamic correlation originates from the instantaneous repulsion between electrons at close range, while static correlation arises when multiple determinants are needed to qualitatively describe the true wavefunction. For example, dynamic effects dominate the correlation energy near the equilibrium geometry of a symmetric diatomic molecule, but static correlation becomes dominant when molecular bonds break into open-shell fragments with long-range correlation effects.

Post-HF wavefunction methods have been developed to capture electron correlation by going beyond the mean-field single-determinant wavefunction approximation.  The method of choice to capture dynamic correlation is Coupled Cluster theory (CC) \cite{kummel2003,bartlett2007,cramer2003, shavitt2009}, whereas Multi-Configurational Self-Consistent Field (MCSCF) theory is one of the primary methods to capture dominant static correlation. \cite{wolinski1987,mcdouall1988,nakano1993,roos1980} Other approaches to capture static correlation are the Density Matrix Renormalization Group (DMRG) method, \cite{white1999,nishino1995,marti2010,stoudenmire2012} full configuration interaction Monte-Carlo (FCIQMC) approaches, \cite{booth2009fermion} or the more recent paired-electron geminal inspired approaches.\cite{limacher:2013,tecmer:2022}  Despite their inherent differences in approach, these methods share the common feature of incorporating electron correlation through a well-defined physical mechanism, for example particle-hole excitations in CC theory or electron-pair excitation operators in geminal theory. In any event, post-HF methods produce more accurate wavefunctions than HF, although the added accuracy comes at the price of a greater computational cost.

We propose a new application of the Generator Coordinate Method (GCM) \cite{hill1954, griffin1957} as a means to go beyond the mean-field approximation in quantum chemistry. The defining feature of the GCM is that it expands the many-body wavefunction over a continuous manifold of reference states labeled by a so-called generator coordinate. Correlations between these reference configurations are then captured via the Hill--Wheeler (HW) equations, a continuous reformulation of the Schr\"odinger equation (see Section \ref{subsection:gcm}).\cite{hill1954,griffin1957}  Thanks to the general underlying principles, the GCM has found widespread use in physics.\cite{ring2004,wong1975,chattopadhyay1978,johansson1978} In quantum chemistry, the GCM formalism was used to develop high-quality HF and Dirac--Fock single-particle basis functions.\cite{mohallem1986,dasilva1989,jorge1996} Expressions similar to the GCM ansatz (Eq.~\eqref{eq:WFN}) have also been used by Alon et al.\cite{alon2005} to symmetrize many-body wavefunctions, and by Pan et al. \cite{pan2004} to extend the variational space during variational calculations. Under the name of the integral transform method, the GCM has been employed to optimize Slater-type one-electron \cite{somojai:1968} and Hylleraas-type two-electron wavefunctions \cite{thakkar:1977} for high-precision calculations in few-electron problems.    

In this work, we envision the GCM in conjunction with a spontaneously broken symmetry for applications in strongly correlated systems. This process is well documented in nuclear structure physics, where strong collective shape correlations have been associated with the spontaneous breaking of the nucleus’ spherical symmetry. \cite{bender:2003, bohr:1998} The key feature of the nuclear GCM is that the continuous manifold of states consists of constrained-HF states, in which the constraint is imposed on the expectation value of the relevant multipole deformation mode. \cite{ring2004} 
The benefits of this approach are twofold. First, the constrained-HF approach generates a potential energy surface (PES) as a function of the constrained multipole moment, with the global minimum of that PES then revealing whether the symmetry is spontaneously broken or not.  Second, the GCM adds correlations around this minimum via the HW equations. As the reference states are generated at the Hartree--Fock level, the cost associated with constructing the manifold of reference states has mean-field scaling.

Spontaneous symmetry breaking is universal in physics and chemistry. In general, symmetry breaking occurs when the lowest-energy solution does not conserve all the symmetry operations that leave the Hamiltonian invariant. \cite{arodz2003, cornell1989} A well-known example of a geometric symmetry breaking process is the Jahn--Teller effect, \cite{cornell1989} whereas an example of a non-geometric symmetry breaking is the particle number symmetry in the Bardeen--Cooper--Schrieffer superconductive state. \cite{bardeen1957theory, phillips2003advanced} The universality of the symmetry breaking process is emphasized by the existence of a second-order phase transition, in which the associated order parameter is provided by the observable quantifying the (broken) symmetry. These phase transitions are sharp for extensive systems, however they are also signaled at the mean-field level for finite-size systems. \cite{ring2004, richter2017quantum, cejnar2010quantum}

In quantum chemistry, symmetry breaking is well understood in the case of restricted HF (RHF) vs unrestricted HF (UHF). The difference between the two approaches is that RHF restricts the molecular spatial orbitals for both spin-up and spin-down sectors to be the same, whereas UHF allows for the optimization over different spatial orbitals in both sectors, leading to a better (but not necessarily \emph{strictly} better) description of the HF energy. Indeed, whereas RHF is incapable of reproducing the PES of closed-shell systems in the stretched bond regime, \cite{parr1950} UHF can successfully exploit the extra variational freedom to produce more accurate energy predictions with respect to the exact full configuration interaction (FCI) result. The trade-off is that we must accept a spin symmetry-broken wavefunction, leading to a reduction in the accuracy for the expectation values of observables other than the energy, such as the magnitude of the total spin $\expval*{\hat{S}^2}$. This choice is known as L\"owdin’s symmetry dilemma, stating that relaxing particular physical symmetry constraints provides better energies through the variational principle, however at the cost of a broken-symmetry wavefunction and associated quantum numbers. \cite{lykos1963discussion}   

\begin{figure}[htb]
\centering
\begin{subfigure}{.5\textwidth}
  \centering
  \includegraphics[scale=1.0]{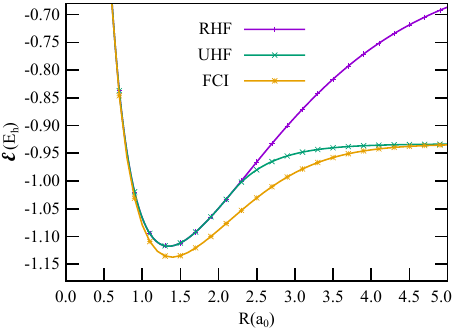}
  \caption{}
  \label{fig:RHFvsUHFenergy}
\end{subfigure}%
\begin{subfigure}{.5\textwidth}
  \centering
  \includegraphics[scale=1.0]{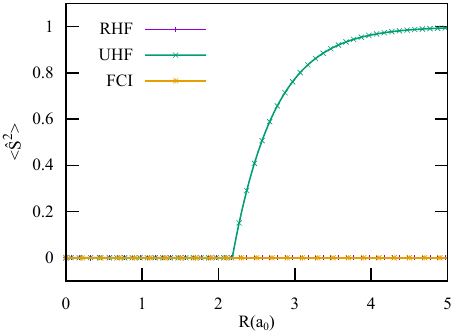}
  \caption{}
  \label{fig:RHFvsUHFspin}
\end{subfigure}
\caption{Total energy $\epsilon$ (a) and $\expval*{\hat{S}^{2}}$ expectation value (b) of the RHF and UHF  ground states for the \ce{H2}/STO-3G binding curve. The $\expval*{\hat{S}^{2}}$ for RHF and FCI are zero across the bond length.}
\label{fig:h2_cfp}
\end{figure}

The prototypical example of mean-field symmetry breaking in the \ce{H2} molecule in a minimal basis set is illustrated in Fig.~\ref{fig:h2_cfp}.  As can be seen in the left panel, UHF provides no variational advantage over RHF near the equilibrium bond distance.  In contrast, the UHF state provides a strictly better energy with respect to RHF as soon as the bond is sufficiently stretched. For the UHF wavefunction, the spin-up electron (or $\alpha$ electron) and spin-down electron (or $\beta$ electron) occupy different molecular orbitals which are both localized around opposite hydrogen atoms.  As a result, the UHF solution is no longer an eigenstate of the spin operator, introducing so-called spin contamination \cite{cassamchenai:1993,cassamchenai:1998,cassam2015spin} in the wavefunction, as can be observed in Fig.~\ref{fig:h2_cfp}\textcolor{blue}{b} where the expectation value $\expval*{\hat{S}^2}$ is plotted as a function of the bond distance. At this point, the symmetry-breaking interpretation of the UHF protocol becomes apparent.  Although the Hamiltonian is spin symmetric, the resulting UHF state is not, as swapping the $\alpha$ and $\beta$ spin functions creates a different UHF state, which is however energetically degenerate with the original UHF.  We will refer to this second \textoverline{UHF} state as the spin-swapped or dual UHF state further on, as it is important for spin-symmetry restoration purposes.  Note that the dual of the \textoverline{UHF} state leads back to the original UHF, and that the dual of the RHF state is the  RHF itself.  
The transition point that marks the symmetry breaking is known as the Coulson--Fischer (CF) point.\cite{coulson1949}  The exact location of the CF point depends on the basis set and the system. In some cases, such as in the \ce{O2^{2+}} dication \cite{nobes1991} and \ce{F2} molecule,\cite{purwanto2008} the CF point can even occur before the equilibrium geometry, leading to a qualitatively incorrect unbound UHF potential energy surface.  Furthermore, this symmetry breaking can also be induced by changes in the nuclear charge\cite{burton2021critical} or externally applied electric fields,\cite{cunha2021} and is particularly prevalent in higher-energy HF solutions.\cite{burtonwales}  

L\"owdin’s symmetry dilemma can be overcome by combining variational approaches with projection methods on the desired (broken) quantum numbers.  Many flavours have been proposed in the literature, mostly within the context of Hartree-Fock theory, such as the extended Hartree-Fock method\cite{lowdin:1960,biskupic:1982,schlegel:1986} or projected Hartree-Fock\cite{jimenez2012} method.  More recently, Thom and coworkers\cite{thom2009hartree} proposed a Non-Orthogonal Configuration Interaction (NOCI) approach in which the basis configurations consist of the RHF and UHF state, augmented with the \textoverline{UHF} to allow for proper spin projections  
%
%
%
 \begin{equation} 
 \ket{\textrm{NOCI}} = C_{\textrm{RHF}} \ket{\textrm{RHF}} + C_{\textrm{UHF}} \ket{\textrm{UHF}} + C_{\overline{\textrm{UHF}}} \ket{\overline{\textrm{UHF}}}.
 \label{eq:hHF}
 \end{equation}
Several key aspects of this approach can be highlighted.   First, the $\ket{\textrm{RHF}}$, $\ket{\textrm{UHF}}$ and $\ket{\overline{\textrm{UHF}}}$ basis states are not necessarily orthogonal, giving rise to a 3-state NOCI approach. Second, due to the presence of the spin-symmetry-broken $\ket{\textrm{UHF}}$ and $\ket{\overline{\textrm{UHF}}}$ states, the NOCI state (\ref{eq:hHF}) does not automatically restore the full spin symmetry in general.  Nevertheless, it can be shown (see Appendix \ref{section:appendix:timereversal}) that due to the degeneracy the contributions of both UHF states are the same up to a phase $|C_{\textrm{UHF}}|=|C_{\overline{\textrm{UHF}}}|$. 
This result is reminiscent of the Half-Projected HF method, in which extra correlation is introduced on top of the UHF state by constructing
 \begin{equation} 
 \ket{\textrm{HPHF}} =  C_{\textrm{UHF}} [\ket{\textrm{UHF}} + (-)^{N} \ket{\overline{\textrm{UHF}}}],
 \label{eq:HPHF}
 \end{equation}
for a $2N$ particle state and $S$=0, \cite{smeyers1973, smeyers:1974,smeyers1978half,cox1976,ruiz:2022} which can be regarded as special case of the NOCI state (\ref{eq:hHF}).    
Third, the wavefunction \eqref{eq:hHF} does not provide any benefits over RHF at shorter bond lengths than the CF point as the UHF and RHF solutions coincide and lead to a singular overlap matrix in the NOCI formalism and a corresponding discontinuity in the PES at the CF point.\cite{hiscock2014}  The holomorphic HF (h-HF) approach remedies this artefact by introducing a complex-analytic continuation of the conventional HF ensuring the existence of multiple unique stationary points for all geometries. \cite{hiscock2014,burton2016,burton2018,burton2020} Where real HF solutions disappear, the corresponding h-HF solutions continue to exist with complex-valued orbital coefficients, providing a non-singular overlap matrix and a continuous NOCI basis.\cite{burton2016,burton2019} 
Fourth, while using h-HF solutions as a basis for NOCI can successfully capture electron correlation, this formalism still relies on the existence of solutions that spontaneously break symmetries for at least one geometry of interest. This restriction limits the method’s black-box applicability. 

The purpose of the present work is to generalize the state (\ref{eq:hHF}) towards a spin-GCM procedure.  In this approach, the NOCI (or HW) state employs a manifold of explicitly symmetry-broken solutions from spin-constrained UHF (c-UHF) calculations \cite{andrews:1991,de2023capturing,de2023spin} with the constraint applied to the expectation value of the total spin operator $\langle\hat{S}^2\rangle=S(S+1)$, which defines the effective spin value $S=\sqrt{\frac{1}{4}+\langle\hat{S}^2\rangle}-\frac{1}{2}$.  For each molecular geometry, multiple c-UHF calculations with different spin values are generated to define the basis for the NOCI/HW equations.  In this paper, we explore different combinations of c-UHF solutions for the NOCI basis, which we describe in detail in sections \ref{sec:noci2}, \ref{sec:noci3}, and \ref{sec:nocin}.
The advantage of the spin-GCM is that it does not depend on the existence of spontaneously symmetry-broken solutions to the HF equations, or their holomorphic counterparts, and allows for more flexibility in the choice of NOCI basis states through the spin value constraints in the c-UHF solutions.

The paper is organized as follows.  For completeness, we briefly reintroduce the GCM approach and its relationship to NOCI in Sections~\ref{subsection:gcm} and \ref{subsection:noci}.  The spin-constrained UHF method is discussed in Section~\ref{subsection:cUHF} as a means to generate the manifold of continuous states. The computational procedure is detailed in Section~\ref{subsection:compdetail}, and results for 2-electron systems that are known to be dominated by static (\ce{H2}) or dynamic (\ce{HeH+}) correlation are presented in Section~\ref{sec:results}.   We focus our applications on the bond-breaking  of two-electron diatomics and refer to Ref.~\onlinecite{de2023spin} for a detailed mathematical analysis of the 2-site Hubbard model. A different application of the spin-GCM based on constrained generalized HF (c-GHF) in the context of frustrated spin systems is discussed in an earlier publication by some of us.\cite{de2023capturing}


\section{Methods}
\subsection{Generator Coordinate Method}\label{subsection:gcm}

The GCM assumes the existence of a manifold of wavefunctions $\ket{\psi(a)}$ parameterized by some continuous variable $a$, known as the generator coordinate, which supports a continuous wavefunction expansion  \cite{ring2004}
\begin{equation} 
\ket{\Psi} = \int \dd a\,f(a)\ket{\psi(a)},
\label{eq:WFN}
\end{equation}
where $f(a)$ are the expansion coefficients. Variation of the expectation value of the many-body Hamiltonian with respect to the weight functions $f$
\begin{equation}
\frac{\delta}{\delta f} \left[ \frac{\mel{\Psi}{\hat{H}}{\Psi}}{\braket{\Psi}{\Psi}} \right] = 0,
\label{eq:dHdf}
\end{equation}
leads to the Hill--Wheeler (HW) equation
\begin{equation} 
\int \dd b\,H(a,b)f(b)\ = E\int \dd b\,O(a,b)f(b),
\label{eq:HW}
\end{equation} 
with the Hamiltonian and overlap kernels defined as
\begin{subequations} 
\begin{align}
H(a,b) &= \mel{\psi(a)}{\hat{H}}{\psi(b)},
\\
O(a,b) &= \braket{\psi(a)}{\psi(b)}.
\end{align}
\end{subequations}
The overlap kernel will be denoted as $O$ to avoid confusion with the spin quantum number $S$. At this point, there is no explicit reference to the particular generator coordinate $a$, nor its associated wavefunctions $\ket{\psi(a)}$. 

In the context of the GCM, two primary approaches can be used to identify the reference wavefunctions. The first one aligns with conventional variational approaches and employs an explicit parametrization of the wavefunction $\ket{\psi(a)}$ in terms of the variational parameter $a$, as used in the integral transform method. \cite{somojai:1968, thakkar:1977} The second approach uses an implicit parametrization in which the reference wavefunctions $\ket{\psi(a)}$ are precomputed from a different theory. Here, it is crucial to choose a sufficiently facile category of reference wavefunctions that does not overload the computation of the Hamiltonian and overlap kernel in the HW equations, motivating the choice for Slater determinant wavefunctions from constrained Hartree-Fock calculations\cite{ring2004,bender:2003}. 

\subsection{Non-Orthogonal Configuration Interaction}\label{subsection:noci}

The HW equations in Eq.~\eqref{eq:HW} are defined on a continuous manifold. Solving this continuous integration is impossible from a practical computational point of view, so one usually discretizes the integral over a discrete set of points $a_{i} \in \{a\}$ \cite{ring2004}
\begin{equation}
\sum_{j} H(a_i, a_j) f(a_j) = E \sum_{j} O(a_i, a_j) f(a_j),
\label{eq:discr}
\end{equation}
reducing the HW equations to a generalized eigenvalue problem. In the case of Slater determinant (SD) reference wavefunctions, the discretized HW equations are equivalent to the NOCI secular equations, and we will use both names HW/NOCI interchangeably in this paper. The NOCI wavefunction ($\ket{\Psi_{i}}$) is composed of linear combination of SDs ($\ket{\psi_{i}}$) as
\begin{equation}
\ket{\Psi_{i}} = \sum_{i} c_{i} \ket{\psi_{i}},
\label{eq:linSD}
\end{equation}
with the coefficient $c_{i}$ for the $i$th SD $\ket{\psi_{i}}$. In contrast to conventional configuration interaction expansions, the orbitals between different SDs are not necessarily orthogonal, \cite{thom2009hartree} leading to a non-diagonal overlap matrix between the different SDs
\begin{equation}
\braket{\psi_{i}}{\psi_{k}} = O_{ij} \neq \delta_{ij}.
\label{eq:nonOrt}
\end{equation}
It is important to choose a sufficiently compact and diverse set of NOCI basis states $|\psi_k\rangle$. Compactness is desired from a computational scaling point of view as the generalized eigenvalue problem scales cubically with the number of input states. Diversity matters for computational stability as incorporating SDs with high overlaps ($\braket{\psi_{i}}{\psi_{k}} \approx1$ for $i \neq j$) leads to singular overlap matrices and potential instabilities in solving the generalized eigenvalue equations.\cite{werner1985second, sundstrom2014non} 

While NOCI calculations are mostly adapted to capture static correlation effects, dynamic correlation can be incorporated using second-order perturbation theory. For example, the NOCI-PT2 approach computes a perturbative correction by expanding the first-order wavefunction using the combined set of singly- and doubly-excited configurations obtained from each reference determinant, which defines the first-order interacting space.\cite{burton2020reaching} This approach is analogous to second-order CASPT2 correction applied to CASSCF wavefunctions.\cite{Andersson1990,Andersson1992}

\subsection{Constrained-HF}\label{subsection:cUHF}


Thus far, the GCM has been defined without explicit reference to the generator coordinate $a$.   The idea for the spin-GCM is to use the total spin as the generator coordinate and the associated spin-constrained unrestricted Hartree-Fock (c-UHF) state as reference wavefunctions.\cite{andrews:1991,de2023capturing,de2023spin} Because of the variational nature of HF theory, it is straightforward to incorporate constraints by means of Lagrange multipliers ($\lambda$) leading to an optimization problem of the constrained Hamiltonian
\begin{equation} 
\hat{\mathcal{H}} = \hat{H} + \lambda (\hat{S}^{2} - S(S+1)). 
\label{eq:Lag}
\end{equation} 
Minimizing the expectation value of this constrained Hamiltonian $\langle \mathcal{\hat{H}}\rangle$ will minimize the energy of the unconstrained (Coulomb) Hamiltonian $\expval{\hat{H}}$ subject to the optimization condition
\begin{equation} 
\frac{\partial}{\partial \lambda} \mel{\textrm{c-UHF}(S)}{\hat{\mathcal{H}}}{\textrm{c-UHF}(S)}
= \mel{\textrm{c-UHF}(S)}{\hat{S}^{2}}{\textrm{c-UHF}(S)}- S(S+1) = 0,
\label{eq:constr}
\end{equation} 
in which $S$ is the constrained effective spin value, chosen by the user from $[0, S_\text{max}]$, which for two-electron systems $S_\text{max} = 1$.
The electronic energy of the resulting c-UHF state is then
\begin{equation} 
 \mathcal{E}(S) = \mel{\textrm{c-UHF}(S)}{\hat{H}}{\textrm{c-UHF}(S)},
\label{eq:lagexpener}
\end{equation} 
which becomes an implicit function of the constrained spin value $S$.  The constrained UHF (c-UHF) approach has been introduced in quantum chemistry by Mukherjee and Karplus \cite{mukherjee:1963} to match the dipole moments from UHF computations to the experimentally obtained values, and was subsequently extended towards general one- and two-body observables,\cite{bjorna:1971} such as the total spin.\cite{andrews:1991}  Meanwhile, the method has found particular traction in the field of nuclear structure physics,\cite{barranger:1961,bonche:1990} mostly in context of the GCM.\cite{bender:2003}  

With the c-UHF states now defined, the explicit spin-GCM state becomes
\begin{equation} 
\ket{\Psi} = \int \dd S\,f(S)\ket{\textrm{c-UHF}(S)}
\label{eq:spingcmstate}
\end{equation}
in which each $\ket{\textrm{c-UHF}(S)}$ is obtained through a c-UHF computation with spin constrained to the value $S$. It is worth mentioning that c-UHF has limitations, as its convergence is not always guaranteed. In the remainder of the paper, we will omit the $S$ functional notation in
$\ket{\textrm{c-UHF}} \equiv \ket{\textrm{c-UHF}(S)}$ and $\mathcal{E} = \mathcal{E}(S)$ for simplicity, unless the context is not  clear. 
The generator coordinate can be multivariate in general, however there is only one single constraint on $S$ in this work. Therefore, we will impose the convergence condition (\ref{eq:constr}) using a line search in $\lambda$, in combination with the conventional Roothaan--Hall self-consistent-field procedure. Expressions for the Fock matrix of the constrained Hamiltonian \eqref{eq:Lag} can be found in Appendix \ref{section:appendix:timereversal}. 

For each system, we can identify two special values of $S$. The $S = 0$ solution coincides with the RHF solution, whereas the $S = S_{\textrm{UHF}}$ is equivalent to the conventional UHF solution. The extra feature of the latter solution is that it not only minimizes the constrained Hamiltonian for the specific spin value $S = S_{\textrm{UHF}}$, but also produces the minimal energy for \emph{all} possible constrained spin values
\begin{equation} 
\mathcal{E}(S_{\textrm{UHF}}) \leq \mathcal{E}(S), \quad \forall S,
\label{eq:coroll}
\end{equation} 
by virtue of the variational principle. A corollary is that the ground state energy of the spin-GCM will always be strictly lower than the UHF prediction. Similarly, if at least the $\ket{\textrm{c-UHF}(S=0)}$ and $\ket{\textrm{c-UHF}(S=S_{\textrm{UHF}})}$ are included in the NOCI, the spin-GCM energy will also be lower than (or equal to) the h-HF-based NOCI in the symmetry broken regime. 

\subsection{Computational Details}\label{subsection:compdetail}

With the essential theoretical aspects of the GCM (Section \ref{subsection:gcm}), NOCI (Section \ref{subsection:noci}), and c-UHF (Section \ref{subsection:cUHF}) now defined, we now specify the computational procedure. All c-UHF computations have been implemented in an in-house Python implementation,\cite{qunbrepo} and the NOCI and PT2 corrections were done with a modified Q-Chem package\cite{shao2015advances} using the LibGNME library to compute the required nonorthogonal matrix elements.\cite{BurtonLibGNME1,BurtonLibGNME2,libgnme} c-UHF and NOCI extensions have also been implemented in the GQCP software package.\cite{lemmens2021gqcp}  Each computation is performed for a specified geometry, which reduces to just an interatomic bond distance $R$ for the dimers considered in the present work. The computational procedure is as follows:


\begin{enumerate}
\item{Define a discrete set of spin values $S_{i}$ $(i = 1,\dots,n)$. We will refer to this set of spin values as the grid. Multiple strategies are possible to choose a proper grid reflecting a balanced trade-off between completeness and compactness of the generated reference states in the NOCI. }
\item{For each imposed spin value $S_{i}$, a c-UHF computation is performed via the procedure outlined in Section \ref{subsection:cUHF}. The resulting $\ket{\textrm{c-UHF($S_{i}$)}}$ state is added to the set of HW/NOCI basis states. This step is repeated for all $n$ spin values. At this point, the $\mathcal{E}(S)$ PES will provide prior information on which $\ket{\textrm{c-UHF(S)}}$ states to prioritize in the HW/NOCI expansion. For instance, the conventional $\ket{\textrm{UHF}} = \ket{\textrm{c-UHF($S_{\textrm{UHF}}$)}}$ state will always emerge as the global minimum of $\mathcal{E}(S)$, and will therefore be an important state to include in the HW/NOCI.}
\item{Compile all computed $\ket{\textrm{c-UHF($S_{i}$)}}$ states, compute the Hamiltonian and overlap kernels, and solve the HW/NOCI equations to obtain a prediction of the ground-state energy.  The separation of the null space from the relevant subspace of linearly independent basis states for the GCM was achieved by prediagonalizing the NOCI overlap matrix and then projecting into the subspace of non-zero eigenvalues.  The threshold for discarded states was set to the standard $10^{-8}$ in LibGNME.}  
\end{enumerate}
The resulting eigenstate $\ket{\Psi}$ (Eq.~\eqref{eq:linSD}) will not necessarily preserve spin symmetry. For that to happen, one must also apply spin projection\cite{jimenez2012, mayer1980} to the c-UHF basis states, in a similar fashion to the localized spin rotations proposed in Ref.~\onlinecite{Lee2022}.
We will only discuss two-electron systems here, for which the addition of the spin-swapped or dual $\ket{\overline{\textrm{c-UHF}}}$ state to the HW/NOCI set will automatically restore the spin symmetry, see Appendix \ref{section:appendix:timereversal}. 

\section{Results and Discussion}
\label{sec:results}

We have selected the \ce{H2} and \ce{HeH+} dimers in various atomic basis sets as simple representatives of molecules dominated by static and dynamic correlation, respectively. The \ce{HeH+} dimer is particularly interesting as there is no spontaneously broken spin symmetry observed in the UHF solution over the whole dissociation curve as the system is dominated by dynamic correlation. Dynamic correlation also becomes more important in \ce{H2} as the basis set is improved from minimal (STO-3G) to double-zeta (cc-pVDZ). These tests allow us to gauge the strengths and weaknesses of the spin-GCM.

\subsection{\ce{H2}/STO-3G: the case of static correlation}

We commence our investigations with \ce{H2} in the STO-3G basis, which is known to be dominated by static correlation in the dissociation limit.  A more formal mathematical study of the related 2-site Hubbard model can be found in Ref.\ \onlinecite{de2023spin}.

\subsubsection{c-UHF phase diagram of \ce{H2}/STO-3G}
We first investigate the c-UHF solutions by considering the unrestricted Hartree-Fock (UHF) solution of the constrained Hamiltonian $\hat{\mathcal{H}}$ (Eq.~\eqref{eq:Lag}) as a function of the Lagrange multiplier $\lambda$.  An important existence question for the c-UHF formalism is whether the constrained Hamiltonian $\hat{\mathcal{H}}$ of eq.~(\ref{eq:Lag}) can support a UHF Slater determinant wavefunction with a preset spin expectation value $S(S+1)$ for at least one Lagrange multiplier $\lambda$.  For this, we reconsider the constrained Hamiltonian eq.~(\ref{eq:Lag}), however with $\lambda$ as a free input parameter
\begin{equation}\label{eq:Hlambda}
    \hat{\mathcal{H}}(\lambda)=\hat{H} +\lambda \hat{S}^2.
\end{equation}
We will omit the constant term $-\lambda S(S+1)$ in the definition of $\hat{\mathcal{H}}(\lambda)$ as it does not affect the structure of the wavefunction.  

\begin{figure}[htb]
\centering
\begin{subfigure}{.5\textwidth}
  \centering
  \includegraphics[scale=1.0]{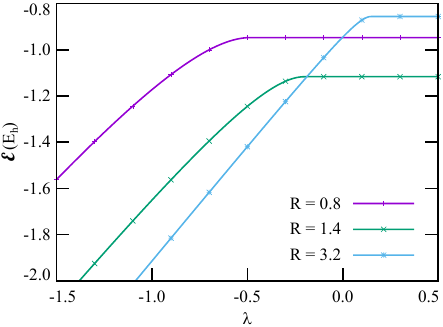}
  \caption{}
  \label{fig:PTenergy}
\end{subfigure}%
\begin{subfigure}{.5\textwidth}
  \centering
  \includegraphics[scale=1.0]{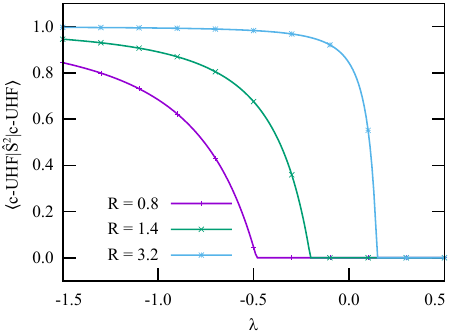}
  \caption{}
  \label{fig:PTspin}
\end{subfigure}
\caption{a) Energy $\epsilon$ and b) the expectation value of total spin as a function of $\lambda$ for $R = 1.4\,\mathrm{a_0}$ (equilibrium), for \ce{H2}/STO-3G. All bond lengths are given in atomic units ($a_0$)}
\label{fig:h2_631g}
\end{figure}

The Hamiltonian $\hat{\mathcal{H}}(\lambda)$ can be interpreted as a competition between the regular Coulomb Hamiltonian $\hat{H}$ and the total-spin operator $\lambda\hat{S}^{2}$ depending on the value of $\lambda$.  Whenever $\lambda$ is sufficiently large and positive, the UHF solution will be equivalent to the RHF solution as any non-zero spin contribution will be penalized. In the opposite direction, for $\lambda$ sufficiently large and negative, the total spin operator will aim to maximize the total spin or spin contamination towards a value of $S_{\textrm{max}}(S_{\textrm{max}}+1)=1$, regardless of the Coulomb interactions.  
Therefore, the Lagrange multiplier $\lambda$ can be used to \emph{enforce} spin symmetry breaking and to create states with a desired spin expectation value.  This is illustrated in Fig.~\ref{fig:h2_631g}, where $\langle\textrm{UHF}|\hat{H}|\textrm{UHF}\rangle$ and $\langle\textrm{UHF}|\hat{S}^2|\textrm{UHF}\rangle$ are plotted as a function of $\lambda$. 
In the context of the c-UHF, we add the constant linear term $-\lambda S(S+1)$ to eq.~(\ref{eq:Hlambda}) to obtain the constrained Hamiltonian eq.~(\ref{eq:Lag}) such that the spin constraint is satisfied when $\mel{\textrm{c-UHF}(S)}{\hat{\mathcal{H}}}{\textrm{c-UHF}(S)}$ is stationary with respect to $\lambda$.
Therefore, the key message from the c-UHF is that the Lagrange multiplier allows for the construction of symmetry-broken states for all possible bond distances, including those that were symmetry conserving in the pure Coulomb regime ($\lambda=0$).  This defining feature sets the spin-GCM apart from the h-HF as the latter may end up in situations where no symmetry-broken states can be identified as seeds for the holomorphic HF states in Eq.\ (\ref{eq:hHF}). 

\begin{figure}[htb]
\centering
\includegraphics[width=0.7\linewidth]{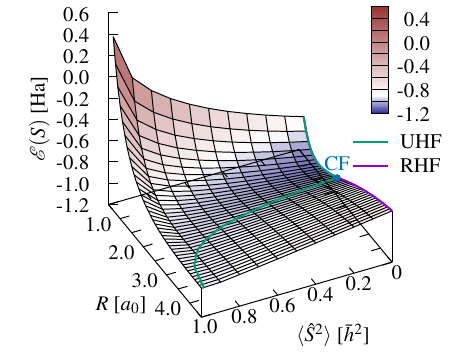}
  \caption{Hamiltonian expectation value $\mathcal{E}(S)$ for \ce{H2}/STO-3G system at different bond lengths $R$. The UHF and RHF solutions are marked by green and purple solid lines on the surface. The Coulson--Fischer point is marked by a blue dot (CF).}
  \label{fig:h2_sto3g_pes}
\end{figure}

The c-UHF energies $\mathcal{E}(S)$ for \ce{H2}/STO-3G are presented in Fig.~\ref{fig:h2_sto3g_pes} as a function of bond distance $R$ and constrained spin expectation value $\langle\textrm{c-UHF}|\hat{S}^2|\textrm{c-UHF}\rangle$. 
A couple of qualitative conclusions can be drawn from Fig.~\ref{fig:h2_sto3g_pes}. For bond distances shorter than the Coulson--Fischer (CF) point, the minimum of the $\mathcal{E}(S)$ cross section is found at $S = 0$, whereas the minimum can be found at a non-zero $S$ beyond the CF point, in line with the conventional UHF solution. The minimum $\mathcal{E}(S_{\textrm{UHF}})$ for each bond distance, as per Eq.~\eqref{eq:coroll}, is marked by the UHF curve in Fig.~\ref{fig:h2_sto3g_pes}. The UHF and RHF PES of Fig.~\ref{fig:h2_cfp}\textcolor{blue}{a} can be reconstructed from a projection of the $S = 0$ (purple RHF curve) and $S = S_{\textrm{UHF}}$ (green UHF curve) solutions, respectively, onto the $\mathcal{E}(S)$ vs $R$ plane, whereas the spin-contamination plots of Fig.~\ref{fig:h2_cfp}\textcolor{blue}{b} are the projection of the respective curves on the $S$ vs $R$ plane. Although the minimum of $\mathcal{E}(S)$ changes qualitatively at the CF point, signaling the symmetry-breaking phase transition, the $\mathcal{E}(S)$ varies smoothly as a function of $R$. This suggests that a slowly varying, or even fixed grid, may be sufficient for the HW/NOCI equations to capture all correlations, as long as the minimum of the $\mathcal{E}(S)$ surface is sufficiently sampled.


\subsubsection{Minimal NOCI: NOCI(2,c-HPHF) and NOCI(2,RHF+c-UHF)} \label{sec:noci2}

\begin{figure}[htb]
\centering
 \includegraphics[scale=1.0]{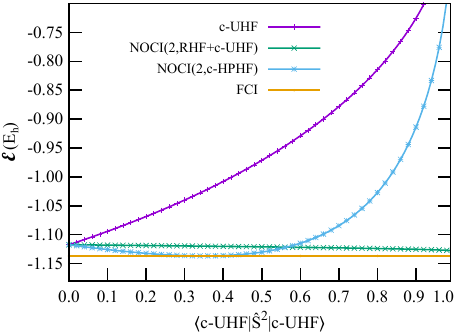}
 \includegraphics[scale=1.0]{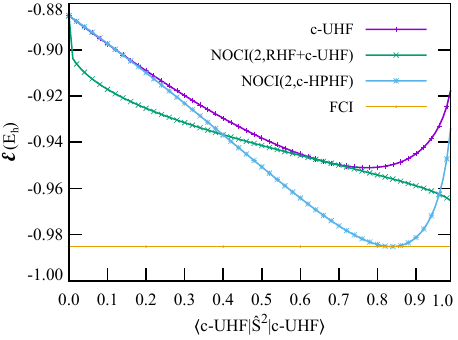}
  \caption{Ground-state energy ($\epsilon$) of \ce{H2}/STO-3G with respect to $\langle\textrm{c-UHF}|\hat{S}^2|\textrm{c-UHF}\rangle$ for c-UHF, NOCI(2,RHF+c-UHF), and NOCI(2,c-HPHF) for (a) $R=1.3459\,\mathrm{a_0}$ and (b) $R=3.0000\,\mathrm{a_0}$. FCI values are shown as full lines to guide the eye and are independent of the spin squared expectation value of the reference determinant.}
  \label{fig:h2_sto3g_noci}
\end{figure}

Once c-UHF states have been determined for each $R$, one can proceed with the spin-GCM procedure. For \ce{H2}/STO-3G, the singlet Hilbert space is 3-dimensional.
For this reason, a HW/NOCI grid with three or more linearly independent grid points should recover the exact energy, in analogy with the 2-site Hubbard model. \cite{de2023spin} We propose the following two approaches that involve only two NOCI basis states:
\begin{enumerate}
\item{HW/NOCI with just the $\ket{\textrm{c-UHF}(S)}$ and its dual $\ket{\overline{\textrm{c-UHF}}(S)}$ at the same $S$. We refer to this as NOCI(2,c-HPHF) as the procedure can be interpreted as a generalization of the HPHF method in Eq.\ \eqref{eq:HPHF}, however with the c-UHF states involved, rather than the UHF.  The ``2'' in the notation emphasizes the fact that only two states are involved.  The spin is naturally restored by construction for 2-electron systems, regardless of the constrained spin value $\langle\textrm{c-UHF}|\hat{S}^2|\textrm{c-UHF}\rangle$ that enters the underlying input c-UHF states (see Appendix \ref{section:appendix:timereversal} ).}
\begin{equation}
|\textrm{NOCI}(2,\textrm{c-HPHF})(S) \rangle = C_{\textrm{c-UHF}}|\textrm{c-UHF}(S)\rangle +C_{\overline{\textrm{c-UHF}}} \ket{\overline{\textrm{c-UHF}}(S)}
\label{eq:HPHF}
\end{equation}

\item{HW/NOCI with just the $\ket{\textrm{c-UHF}(S)}$ at given $S$ and $\ket{\textrm{RHF}} = \ket{\textrm{c-UHF}(0)}$ states. We call this NOCI(2,RHF+c-UHF) because there are again 2 states involved, specified by their respective names. The spin is not restored explicitly because the dual state of $\ket{\textrm{c-UHF}(S)}$ is not included.}
\begin{equation}
|\textrm{NOCI}(2,\textrm{RHF+c-UHF})(S) \rangle = C_{\textrm{RHF}}|\textrm{RHF}\rangle + C_{\textrm{c-UHF}}|\textrm{c-UHF}(S)\rangle
\end{equation}

\end{enumerate}
We present results for both methods in Fig.~\ref{fig:h2_sto3g_noci} for two different different bond lengths, and compare with the underlying input c-UHF computations and the target FCI value.  We will only consider ground state properties in the present work, although excited states can also be considered. 

The total energy is plotted as a function of the constrained spin $\langle\textrm{c-UHF}|\hat{S}^{2}|\textrm{c-UHF}\rangle$ of the underlying c-UHF state for both NOCI(2,c+HPHF) and NOCI(2,RHF+c-UHF). The FCI ground state energy is added for reference. In the case of NOCI(2,RHF+c-UHF), there is no minimum in the energy with respect to the constrained spin $\langle\textrm{c-UHF}|\hat{S}^{2}|\textrm{c-UHF}\rangle$. Compared to c-UHF, the method captures a fair amount of the correlation energy over the entire range of $\langle\textrm{c-UHF}|\hat{S}^{2}|\textrm{c-UHF}\rangle$ but not all. This is not entirely surprising given the lack of proper spin restoration for this method.  We therefore abandon NOCI(2,RHF+c-UHF) in the remainder of this work, and only focus on methods with full spin restoration. 

Remarkably, the NOCI(2,c-HPHF) always gives a minimum for a non-zero $\langle\textrm{c-UHF}|\hat{S}^{2}|\textrm{c-UHF}\rangle$ state, even if the symmetry is not broken at the UHF level (See Fig \ref{fig:h2_sto3g_noci}(a)).  In regimes where UHF is energetically favourable over RHF (See Fig \ref{fig:h2_sto3g_noci}(b), the NOCI(2,c-HPHF) provides a clear energetic advantage over UHF, visually reproducing the FCI target energy at the variational minimum.  
The minimum of NOCI(2,c-HPHF) is equivalent to the HPHF prediction because HPHF is a variation-after-projection approach that minimizes the energy of the state Eq.~\eqref{eq:HPHF}.\cite{smeyers:1974} 

\subsection{\ce{H2}/cc-pVDZ: adding dynamical correlation}

\begin{figure}[b]
\centering
  \includegraphics[scale=1.0]{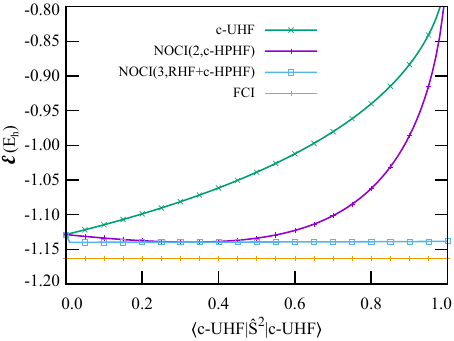}
  \includegraphics[scale=1.0]{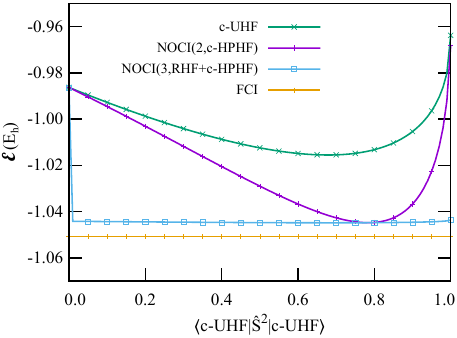}
  \caption{Energy curves for \ce{H2}/cc-pVDZ with respect to $\langle \textrm{c-UHF}|\hat{S}^2|\textrm{c-UHF}\rangle$ for FCI, UHF, NOCI(2,c-HPHF), and NOCI(3,RHF+c-HPHF) methods at $R$ = 1.4 $a_0$ (left) and $R$ = 3.0 $a_0$ (right).}
  \label{fig:h2_ccpvdz_noci}
\end{figure} 

\begin{figure}[htb]
   \centering
    \includegraphics[scale=1.2]{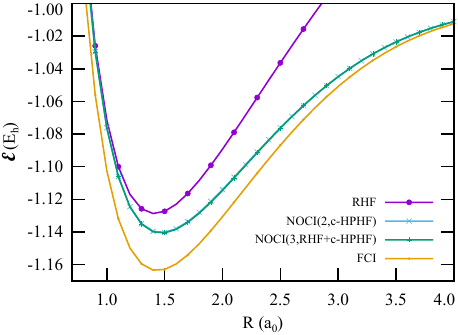}
  \caption{\ce{H2}/cc-pVDZ potential energy curves for RHF, NOCI(2,c-HPHF) and NOCI(3,RHF+c-HPHF) with respect to FCI. The NOCI(2,c-HPHF) and NOCI(3,RHF+c-HPHF) curves are superimposed.}
  \label{fig:noci2hphf:pes}
\end{figure}

Moving to a larger basis set introduces extra dynamical correlation, for instance in the \ce{H2}/cc-pVDZ system.  The c-UHF and NOCI(2,c-HPHF) results are plotted in Fig.~\ref{fig:h2_ccpvdz_noci} for both equilibrium ($R = 1.3794\,\mathrm{a_0}$) and near dissociation ($R = 3.0000\,\mathrm{a_0}$) as a function of constrained spin $\langle\textrm{c-UHF}|\hat{S}^2|\textrm{c-UHF}\rangle$.  
The results closely mirror our observations for the \ce{H2}/STO-3G system, however there are also some significant quantitative differences.  At the equilibrium distance, the NOCI(2,c-HPHF) shows a clear variational minimum with respect to the c-UHF $S^2$ value, however the percentage of correlation energy, defined as 
\begin{equation}
\%E_c=\frac{E_{\textrm{method}}-E_{\textrm{HF}}}{E_{\textrm{FCI}}-E_{\textrm{HF}}}\times100\%.
\end{equation}
amounts to $\% E_c=31.50\,\%$ of the FCI correlation energy.  The situation improves near dissociation where the percentage of correlation energy becomes $\%E_c=90.64\,\%$.  NOCI(2,c-HPHF) results for the full dissociation curve are presented in Fig.~\ref{fig:noci2hphf:pes}.
The results across the potential energy curve are consistent with the proposition that the spin-GCM primarily captures static correlation, but is not sufficiently suitable to include dynamical correlation.   This is confirmed when adding missing correlation energy using NOCI-PT2 corrections \cite{burton2020reaching} on top of NOCI(2,c-HPHF), which improves the percentage of correlation energy to $\%E_c=86.54\,\%$ at equilibrium.

\subsubsection{Increasing the NOCI space: NOCI(3,RHF+c-HPHF)}
\label{sec:noci3}
The question at this point is whether adding more basis states can capture more of the correlation energy.  A natural choice is to add the RHF state, leading to the following, third, NOCI approach
\begin{enumerate}
\setcounter{enumi}{2}
\item{HW/NOCI with the $\ket{\textrm{c-UHF}(S)}$ and dual $\ket{\overline{\textrm{c-UHF}}(S)}$ at the same $S$, augmented with the $\ket{\textrm{RHF}}$ state. We refer to this as NOCI(3,RHF+c-HPHF) as it basically adds the RHF state to the previously defined NOCI(2,c-HPHF) approach. The spin is also restored by construction for a 2-electron system, opposed to the discarded NOCI(2,RHF+c-UHF) approach.}
\begin{equation}
|\textrm{NOCI}(3,\textrm{RHF+c-HPHF})(S) \rangle = C_{\textrm{RHF}}|\textrm{RHF} \rangle + C_{\textrm{c-UHF}} |\textrm{c-UHF}(S)\rangle + C_{\overline{\textrm{c-UHF}}} \ket{\overline{\textrm{c-UHF}}(S)}
\end{equation}
\end{enumerate}
The NOCI(3,RHF+c-HPHF) energies for \ce{H2}/cc-pVDZ with respect to $\langle\textrm{c-UHF}|\hat{S}^{2}|\textrm{c-UHF}\rangle$ are also plotted in Fig.~\ref{fig:h2_ccpvdz_noci}.  A first observation is that the energy profiles are remarkably flat as a function of $\langle\textrm{c-UHF}|\hat{S}^{2}|\textrm{c-UHF}\rangle$.  A second observation is that adding the RHF state only provides a small variational advantage over NOCI(2,c-HPHF).  This observation is confirmed across the full bond dissociation in Fig.~\ref{fig:noci2hphf:pes} where  NOCI(2,c-HPHF) and  NOCI(3,RHF+c-HPHF) are visually very close.  Upon incorporating the NOCI-PT2 calculation in addition to NOCI (3,RHF+c-HPHF), the percentage of captured correlation energy $\%E_c$ increases from $32.23\%$ and $90.65\%$ to $87.06\,\%$ and $99.16\,\%$ respectively at shorter and longer bond lengths. Since the NOCI eigenstate is spin-pure and all the excited determinants in the NOCI-PT2 calculation are present alongside their dual states, the NOCI-PT2 wavefunction correction is also spin-pure.


\subsubsection{NOCI(n): a Generator Coordinate Method}
\label{sec:nocin}

\begin{figure}[!htb]
    \centering
    \includegraphics[width=0.45\linewidth]{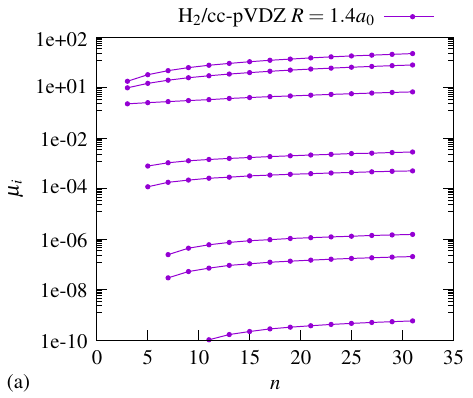}
    \includegraphics[width=0.45\linewidth]{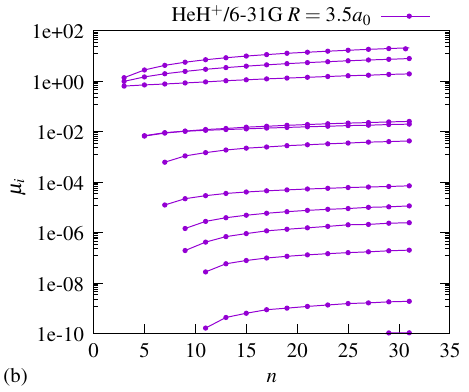}
    \caption{Eigenvalues $\mu_i$ ($i=1,\dots,n$) of the NOCI($n$) overlap matrix for (a) \ce{H2}/cc-pVDZ at $R=1.4 a_0$ and (b) \ce{HeH+}/6-31G at $R=3.5 a_0$. Values not shown fall under the $10^{-10}$ threshold for plotting.}
    \label{fig:nocioverlap-spectrum}
\end{figure}

In a final NOCI approach, we fully embrace the philosophy of the GCM by incorporating a significantly denser sampling of $|\textrm{c-UHF}(S)\rangle$ states on a discrete grid of $S$ values for the NOCI basis in the Hill--Wheeler formalism.  In particular, we propose a HW/NOCI approach on a simple equidistant grid of c-UHF points as a fourth and final method:
\begin{enumerate}
\setcounter{enumi}{3}
\item{HW/NOCI in which the full spin range $S \in [0:S_{\textrm{max}}]$ is partitioned in a grid of $\frac{n}{2}(n+1)$ evenly spaced points $S_{i} \in [0, \frac{2S_{\textrm{max}}}{n-1}, \frac{4S_{\textrm{max}}}{n-1}, ..., S_{\textrm{max}}]$, where $n$ is odd. At each $S_{i}$, both $\ket{\textrm{c-UHF($S_{i}$)}}$ and their dual $\ket{\overline{\textrm{c-UHF}}(S_{i})}$ are included, in order to preserve spin symmetry (See Appendix \ref{section:appendix:timereversal} ).   Note that the $\ket{\textrm{RHF}}=\ket{\textrm{c-UHF}(S_1=0)}$ state is self-dual, so we end up with exactly $n$ c-UHF grid points in the HW/NOCI. We refer to this as NOCI($n$). Given the definition of the dual state, the singlet $S=0$ comes with a positive relative phase $C_{\overline{\textrm{c-UHF}}} = + C_{\textrm{c-UHF}}$, whereas the $S=1$ has a negative phase $C_{\overline{\textrm{c-UHF}}} = - C_{\textrm{c-UHF}}$.  There is no need to explicitly encode these relations in the NOCI procedure as the relative phases are recovered through diagonalization. It would be possible to pre-organize the cUHF and dual basis states  into their respective symmetry sectors to save some computational expense, but we do not employ this approach  given the simplicity of the 2-electron problem.
The resulting wavefunction is defined as 
\begin{equation}
\begin{split}
|\textrm{NOCI}(n) \rangle = & \, C_{\textrm{RHF}}|\textrm{RHF} \rangle  
+
\\
& +C_{\textrm{c-UHF}(\tfrac{2S_{\textrm{max}}}{n-1})}|\textrm{c-UHF}(\tfrac{2S_{\textrm{max}}}{n-1})\rangle + C_{\overline{\textrm{c-UHF}}(\tfrac{2S_{\textrm{max}}}{n-1})} |\overline{\textrm{c-UHF}}(\tfrac{2S_{\textrm{max}}}{n-1})\rangle \\
& +\dots \\
& +C_{\textrm{c-UHF}(S_{\textrm{max}})}|\textrm{c-UHF}(S_{max})\rangle + C_{\overline{\textrm{c-UHF}}(S_{\textrm{max}})} \ket{\overline{\textrm{c-UHF}}(S_{\textrm{max}})}.
\end{split}
\end{equation}}

\end{enumerate}
The hypothesis is that NOCI($n$) will allow for a systematic improvement of the correlation energy as the grid density $n$ is increased.  However, increasing the grid density also implies a growing overlap between neighboring $|\textrm{c-UHF}\rangle$ state, which will eventually lead to an exhaustion of the available linearly independent subspace of the Hilbert space in which the c-UHF resides.  To investigate the size of the available subspace for \ce{H2}/cc-pVDZ, we report the eigenvalue spectrum $\mu_i$ ($i=1,\dots,n$) of the overlap matrix $O$ of NOCI($n$) with increasing $n$ in Fig.~\ref{fig:nocioverlap-spectrum}.  
The Figure presents the spectrum for $R=1.4\,\unit{a_0}$, however other bond distances are qualitatively similar.  We notice that the spectrum is highly structured with a few dominant eigenvalues that remain stable with increasing $n$.  As the numerical threshold for the null space in the LibGNME package was set to $10^{-8}$, approximately 7 linearly independent states are sufficient to span the Hilbert subspace, meaning that NOCI($n$) should effectively converge around $n\sim7$.

\begin{figure}[htb]
\centering
    \includegraphics[scale=1.0]{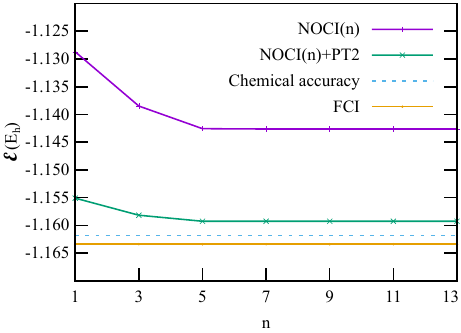}
    \includegraphics[scale=1.0]{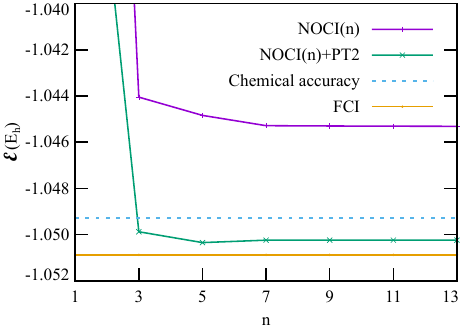}
  \caption{NOCI($n$) and NOCI($n$)+PT2 results for \ce{H2}/cc-pVDZ at (a) $R = 1.40 a_0$ and (b) $R = 3.00 a_0$.  Reference FCI values are given as a full line.}
  \label{fig:noci_convergence}
\end{figure}

\begin{table}[H]
 \caption{NOCI energies ($E_h$) and percentage of correlation energy captured in NOCI(2,RHF+c-UHF), NOCI(2,c-HPHF), NOCI(3,RHF+c-HPHF), NOCI\,(3), NOCI\,(5), NOCI\,(7), NOCI\,(9) and FCI calculations on \ce{H2}/cc-pVDZ.  The dimension of the FCI Hilbert space is 100.}
 \label{table:ccpvdz1st}
\begin{ruledtabular}
 \begin{tabular}{lccccc} 
 & \multicolumn{4}{c}{$R = 1.4 a_0$}  
\\
 \cline{2-6} 
 & $E_\text{method} / \mathrm{E_h}$  & $\%E_c$ & $\langle \hat{S}^2 \rangle$ & $E_\text{method+PT2} / \mathrm{E_h}$  & $\%E_c$ 
\\ \hline
RHF                & -1.12871 & 00.00 & 0.000 & -1.15506 & 75.96 \\
UHF                & -1.12871 & 00.00 & 0.000 & -1.15506 & 75.96 \\
NOCI(2,RHF+c-UHF)  & -1.13400 & 15.25 & 0.013 & -1.15712 & 81.90 \\ 
NOCI(2,c-HPHF)     & -1.13963 & 31.50 & 0.000 & -1.15873 & 86.54 \\ 
NOCI(3,RHF+c-HPHF) & -1.13989 & 32.23 & 0.000 & -1.15891 & 87.06 \\
NOCI\,(3)          & -1.13848 & 28.17 & 0.000 & -1.15817 & 84.93 \\
NOCI\,(5)          & -1.14256 & 39.93 & 0.000 & -1.15925 & 88.04 \\
NOCI\,(7)          & -1.14262 & 40.10 & 0.000 & -1.15926 & 88.07 \\
NOCI\,(9)          & -1.14263 & 40.13 & 0.000 & -1.15926 & 88.07 \\ 
FCI($\textrm{dim}=100$)  & -1.16340 & 100.00 & 0.000 & N/A & N/A
\\ \hline \hline
 & \multicolumn{4}{c}{$R = 3.0 a_0$}
\\
 \cline{2-6} 
& $E_\text{method} / \mathrm{E_h}$ & $\%E_c$ & $\langle \hat{S}^2 \rangle$ & $E_\text{method+PT2} / \mathrm{E_h}$  & $\%E_c$ 
\\ \hline
RHF                & -0.98630 & 00.00 & 0.000 & -1.02364 & 57.80  \\
UHF                & -1.01554 & 45.28 & 0.678  & -1.04351 & 88.56  \\
NOCI(2,RHF+c-UHF)  & -1.02994 & 67.55 & 0.331  & -1.04727 & 94.38  \\
NOCI(2,c-HPHF)     & -1.04483 & 90.64 & 0.000 & -1.05033 & 99.12  \\ 
NOCI(3,RHF+c-HPHF) & -1.04484 & 90.65 & 0.000 & -1.05036 & 99.16  \\
NOCI\,(3)          & -1.04405 & 89.43 & 0.000 & -1.04987 & 98.44  \\
NOCI\,(5)          & -1.04484 & 90.65 & 0.000 & -1.05034 & 99.17  \\
NOCI\,(7)          & -1.04529 & 91.35 & 0.000 & -1.05024 & 99.02  \\
NOCI\,(9)          & -1.04530 & 91.37 & 0.000 & -1.05024 & 99.02  \\ 
FCI($\textrm{dim}=100$)              & -1.05090 & 100.00 & 0.000 & N/A & N/A
 \end{tabular}
\end{ruledtabular}
\end{table}

\begin{figure}[H]
   \centering
    \includegraphics[scale=1.0]{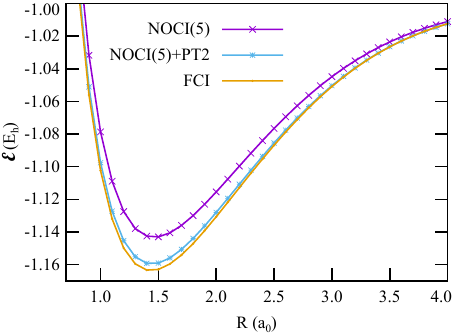}
  \caption{NOCI(5) and NOCI(5)+PT2 PES binding curves for \ce{H2}/cc-pVDZ.}
  \label{fig:nociBindingCurve}
\end{figure}

We present NOCI($n$) results for \ce{H2}/cc-pVDZ at $R = 1.4 a_0$ and $R = 3.0\,a_0$ in Fig.~\ref{fig:noci_convergence} and Table \ref{table:ccpvdz1st}. 
As expected, the NOCI($n$) ground state converges quickly until $n\sim7$, after which no additional correlation energy is gained by increasing the grid density.  
In line with previous results, we see again that only a small fraction of the correlation energy is captured at equilibrium, up to $\%E_c=40.13$ at NOCI(9), whereas a larger portion is achieved beyond dissociation, up to $\%E_c=91.37\%$ for NOCI(9).   While the converged NOCI(9) energy is not within chemical accuracy, introducing dynamic correlation with the NOCI-PT2 correction achieves chemical accuracy for the stretched configuration. It is interesting to note from Table \ref{table:ccpvdz1st} that the NOCI($n$) method does not outperform the limited NOCI(2,c-HPHF) and NOCI(3,RHF+c-HPHF) methods, pointing out that spin projection on a limited space might be comparable to the accurate incorporation of spin fluctuations.  The potential energy curve for NOCI(5) and NOCI(5)+PT2 is shown in Fig.~\ref{fig:nociBindingCurve}, showing a consistent performance across the full range of bond lengths.  

In conclusion, our analysis of the spin-GCM for \ce{H2} in two different basis sets demonstrates that the spin-GCM can capture static correlation in a compact HW/NOCI basis and provides a suitable reference for NOCI-PT2  to capture the remaining dynamical correlation.


\subsection{\ce{HeH+}/6-31G: absence of a spontaneous symmetry breaking}

\begin{figure}[htb]
\centering
\includegraphics[width=0.7\linewidth]{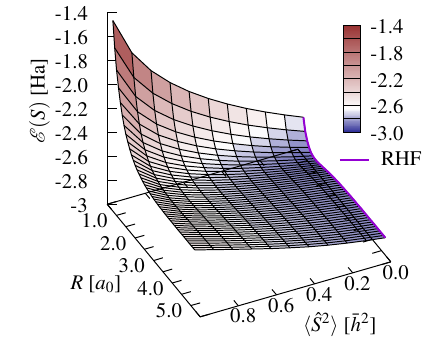}
  \caption{Hamiltonian expectation value $\mathcal{E}(S)$ for \ce{HeH+}/6-31G system at different bond lengths $R$. The RHF solutions are marked by purple solid lines on the surface.}
  \label{fig:heh_631g_pes}
\end{figure}

We now put the spin-GCM method further to the test, by examining its performance on \ce{HeH+}/6-31G, a system with predominantly dynamic correlation across the bond dissociation such that spontaneous spin symmetry breaking is completely absent, making \ce{HeH+} unattainable for h-HF. 
We confirm the absence of the spin-symmetry breaking in Fig.~\ref{fig:heh_631g_pes}, where the c-UHF solutions do not have a non-zero $\langle \textrm{c-UHF}|\hat{S}^2|\textrm{c-UHF}\rangle$ minimum across the entire range of bond lengths.

\begin{figure}[htb]
\centering
    \includegraphics[scale=1.0]{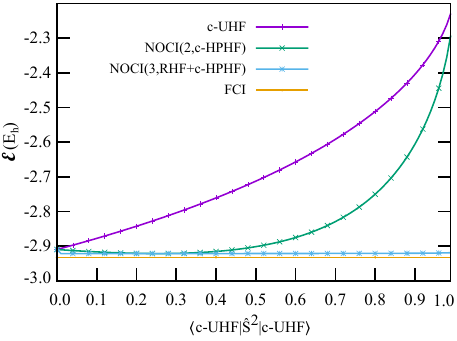}
    \includegraphics[scale=1.0]{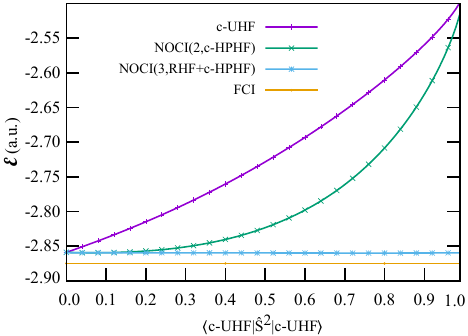} 
  \caption{c-UHF, NOCI(2,c-HPHF) and NOCI(3,RHF+c-HPHF) energies as a function of $\langle\textrm{c-UHF}|\hat{S}^2|\textrm{c-UHF}\rangle$ for \ce{HeH+}/6-31G at (a)$R = 1.5 \unit{a_0}.$ and (b) $R = 3.5 \unit{a_0}$. The FCI reference energy is provided as a full line. }
  \label{fig:heh_6-31g}
\end{figure}

The role of the $\lambda$ Lagrange multiplier in explicitly inducing spin-symmetry-breaking now becomes clear.  The NOCI(2,c-HPHF) does obtain a minimum at non-zero $\langle\textrm{c-UHF}|\hat{S}^2|\textrm{c-UHF}\rangle$ values, capturing $51.92\,\%$ of the correlation energy near equilibrium ($R = 1.50$ $a_0$) but only $3.23\,\%$ near dissociation ($R = 3.50$ $a_0$). The apparent paradox that the spin-GCM is able to better capture correlation in \ce{HeH+} at equilibrium than dissociation is easily resolved by realizing that the dissociated \ce{He} + \ce{H+} is dominated by dynamical correlation on the He atom, which has closed-shell character, whereas some degree of static correlation is encoded in the He-H bond at equilibrium.  Adding the RHF state explicitly does not provide much improvement with a $\%E_c=52.37\%$ and $6.17\%$ at $R = 1.50 a_o$ and $3.50 a_0$ respectively for NOCI(3,RHF+c-HPHF). Detailed numerical results can be found in Table \ref{table:HeH631G1st}. 

\begin{figure}[H]
\centering
\includegraphics[scale=1.0]{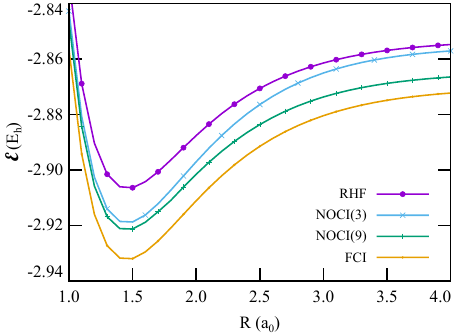}
\includegraphics[scale=1.0]{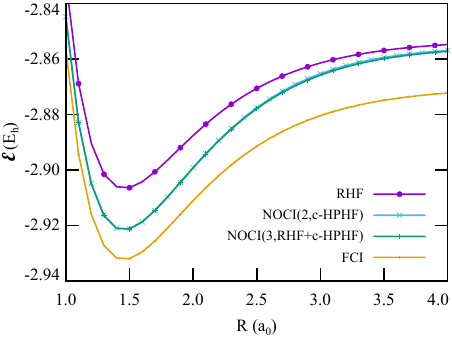}
\caption{The binding curves for NOCI\,(3), and NOCI\,(9) (left) and NOCI\,(2,c-HPHF) and NOCI\,(3,RHF+c-HPHF) (right) compared to the FCI result for \ce{HeH+}/6-31G. The NOCI\,(2,c-HPHF) and NOCI\,(3,RHF+c-HPHF) curves are superimposed.}
\label{fig:heh-6-31g-PES}
\end{figure}

\begin{table}[H]
\caption{NOCI energies ($E_h$) and percentage of correlation energy captured in NOCI(2,RHF+c-UHF), NOCI(2,c-HPHF), NOCI(3,RHF+c-HPHF), NOCI\,(3), NOCI\,(5), NOCI\,(7), NOCI\,(9) and FCI calculations on \ce{HeH+}/6-31G. The dimension of the FCI Hilbert space is 16. }
\label{table:HeH631G1st}
\begin{ruledtabular}
 \begin{tabular}{lccccc} 
 & \multicolumn{4}{c}{$R = 1.50 a_0$}  
\\
 \cline{2-6} 
 & $E_\text{method} / \mathrm{E_h}$  & $\,\%\ \text{Corr.}$  & $\langle \hat{S}^2 \rangle$  & $E_\text{method-PT2} / \mathrm{E_h}$  & $\,\%\ \text{Corr.}$
\\ \hline
RHF=UHF              & -2.90950 & 00.00 & 0.000 & -2.92634 & 74.84 \\  
NOCI(2,RHF+c-UHF)    & -2.91418 & 20.80 & 0.006 & -2.92774 & 81.07\\
NOCI\,(2,c-HPHF)     & -2.92118 & 51.92 & 0.000 & -2.92972 & 89.89 \\
NOCI\,(3,RHF+c-HPHF) & -2.92128 & 52.37 & 0.000 & -2.92976 & 90.07 \\
NOCI\,(3)            & -2.91876 & 41.16 & 0.000 & -2.92930 & 88.02 \\
NOCI\,(5)            & -2.92127 & 52.32 & 0.000 & -2.92980 & 90.25 \\
NOCI\,(7)            & -2.92128 & 52.37 & 0.000 & -2.92978 & 90.16 \\
NOCI\,(9)            & -2.92128 & 52.37 & 0.000 & -2.92979 & 90.20 \\
FCI($\textrm{dim}=16$) & -2.93200 & 100.00 & 0.000 & N/A & N/A  \\ 
\hline \hline
 & \multicolumn{4}{c}{$R = 3.50 a_0$}
\\
 \cline{2-6} 
& $E_\text{method} / \mathrm{E_h}$  & $\,\%\ \text{Corr.}$ & $\langle \hat{S}^2 \rangle$   & $E_\text{method-PT2} / \mathrm{E_h}$  & $\,\%\ \text{Corr.}$ 
\\ \hline
RHF=UHF              & -2.85890 & 0.00  & 0.000 & -2.87076 & 74.13  \\
NOCI(2,RHF+c-UHF)    & -2.85930 & 2.50  & 0.001 & -2.87088 & 74.88  \\
NOCI\,(2,c-HPHF)     & -2.85942 & 3.23  & 0.000 & -2.87110 & 76.32  \\
NOCI\,(3,RHF+c-HPHF) & -2.85989 & 6.17  & 0.000 & -2.87116 & 76.70  \\
NOCI\,(3)            & -2.85957 & 4.17  & 0.000 & -2.87110 & 76.32  \\ 
NOCI\,(5)            & -2.85977 & 5.42  & 0.000 & -2.87114 & 76.57  \\ 
NOCI\,(7)            & -2.86719 & 51.85 & 0.000 & -2.87375 & 92.91 \\ 
NOCI\,(9)            & -2.86879 & 61.87 & 0.000 & -2.87359 & 91.90 \\ 
FCI($\textrm{dim}=16$)  & -2.87490 & 100.00 & 0.000 & N/A & N/A 
 \end{tabular}
\end{ruledtabular}
\end{table}
and the PES are provided in Fig.~\ref{fig:heh-6-31g-PES}.
As expected, the NOCI(2,c-HPHF) and NOCI(3,RHF+c-HPHF) formalisms are not particularly effective themselves for the \ce{HeH+}/6-31G example because the electron correlation in \ce{HeH+} is primarily dynamic in nature. The NOCI-PT2 correction is required to capture the missing dynamic correlation. NOCI(3,RHF+c-HPHF)+PT2 captures $90.07\,\%$ of the missing correlation energy at equilibrium and $76.70\,\%$ is obtained near dissociation, which is a considerable improvement.

\begin{figure}[htb]
 \centering
    \includegraphics[scale=1.0]{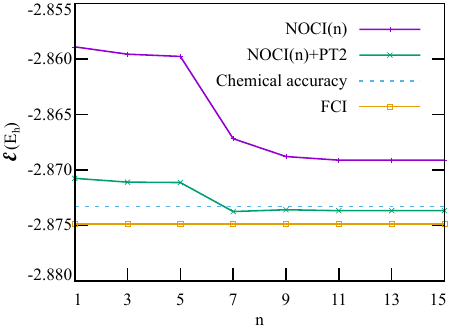}
  \caption{The total energy with respect to $n$ (NOCI dimension) for NOCI\,$(n)$, NOCI\,$(n)$+PT2 calculations compared to the FCI results at $R$ = 3.50 $a_0$ for \ce{HeH+}/6-31G.}
  \label{fig:heh_convergence}
\end{figure}

Finally, NOCI($n$) energies with increasing $n$ near the dissociation bond length are presented in Fig.~\ref{fig:heh_convergence}. 
The eigenvalues of the overlap matrix are presented in Fig.~\ref{fig:nocioverlap-spectrum} (b), pointing out that about 10 linearly independent states are available on the c-UHF grid subspace of the Hilbert space.  We only observe good improvement in capturing missing correlation at the NOCI(7) level in Fig.~\ref{fig:heh_convergence}, pointing out that the static correlation is slightly more intricate than in the \ce{H2}/cc-pVDZ case.  We also observe that NOCI($n$) does not converge within chemical accuracy.  When the PT2 correction is added, chemical accuracy is reached after $n = 7$. These results demonstrate that the spin-GCM approach can be used to build a basis for NOCI even in a dynamically correlated case like \ce{HeH+}, where there is no spontaneous symmetry breaking.

\section{Conclusion}

The generator coordinate method (GCM) is a variational approach to the quantum many-body problem in which interacting many-body wavefunctions are constructed as superpositions of (generally nonorthogonal) eigenstates of a Hamiltonian containing a deformation parameter that breaks a symmetry. For our spin-GCM, the total spin variable is used as the generator coordinate since the Hartree--Fock (HF) mean-field description of molecular dissociation phenomena improves whenever the spin symmetry is explicitly broken from restricted (RHF) to unrestricted HF (UHF). GCM consists of two steps; first, constrained-HF calculations  break the spin symmetry and generate a manifold of states with different values of $S$. Secondly, a Hill-Wheeler or non-orthogonal configuration interaction (HW/NOCI) calculation is performed on the resulting states to restore the broken symmetry and recover the missing correlation between the HF states. 

We have tested this method for 2-electron systems with the STO-3G, 6-31G, and cc-pVDZ basis sets. Our results show that spin-GCM can compute the missing correlation in systems where static correlation is  prominent, and can also  be applied to systems pertaining more to dynamic correlation if used as a precursor for  NOCI-PT2. 

Our spin-GCM extends the symmetry-breaking and restoration framework's versatility beyond the holomorphic HF (h-HF) approaches that rely on the existence of spontaneously symmetry-broken solutions or their holomorphic counterparts.  It will be interesting to extend the spin-GCM to more intricate bond scenarios or more complex systems like transition metal centers, where spin symmetry breaking is pivotal.  Furthermore, the relative importance of the spin-projection over spin-fluctuation deserves further scrutiny in light of extensions towards larger systems.   Lastly, the use of alternative generator coordinates beyond spin creates the potential to address various forms of electron correlation in a wide range of chemical and physical systems.

\section{Acknowledgments}

We thank Xeno De Vriendt and Guillaume Acke for their helpful discussions. We also acknowledge conversations with Alex Thom on the holomorphic Hartree-Fock method, which has been the inspiration for the spin-GCM. We also acknowledge conversations with Patrick Cassam-Chenai about spin-contamination in unrestricted Hartree-Fock approaches. SDB and AA acknowledge the Canada Research Chair program, the CFI, NSERC, and NBIF for financial support. HGAB acknowledges funding from New College, Oxford and Downing College, Cambridge and is a Royal Society University Research Fellow (URF\textbackslash R1\textbackslash 241299) at University College London. Parts of this research were funded by FWO research project G031820N. This research was enabled in part by software provided by the Digital Research Alliance of Canada (alliancecan.ca).

\bibliography{main}
\appendix
\section{Spin and the Coulomb Hamiltonian}\label{section:appendix}
\subsection{Operators}

It is most convenient to express the spin-GCM formalism in second quantized language.  The creation/annihilation operators of an electron in one of $L$ spatial orbital $i$ are given by $\hat{a}_i^\dag$ and $\hat{a}_i$ respectively, dressed with the anticommutation relations
\begin{equation}
    \{\hat{a}_i^\dag,\hat{a}_j\}=\delta_{ij},\quad \{\hat{a}_i^\dag,\hat{a}_j^\dag\}=0,\quad\{\hat{a}_i,\hat{a}_j\}=0,\quad\forall i,j=1,\dots,L.\label{app:anticommutation}
\end{equation}
The orthogonality of the spatial orbitals is implied because of the Kronecker-$\delta$ metric in the anticommutation relations, although a non-orthogonal framework is equally viable, leading to Roothaan equations instead of the conventional Hartree-Fock equations.

We will distinguish between the two different spin sectors by denoting spin-up electrons by their spatial orbital $i$, whereas spin-down electrons will receive a bar $\bar{i}$ notation.  Spin-down electrons obey the same anticommutation relations as the spin-up electrons, whereas spin-up and spin-down electrons anticommute as well mutually
\begin{equation}
    \{\hat{a}_i^\dag,\hat{a}_{\bar{j}}\}=0,\quad \{\hat{a}_i^\dag,\hat{a}_{\bar{j}}^\dag\}=0,\quad\{\hat{a}_i,\hat{a}_{\bar{j}}\}=0,\quad\forall i,j=1,\dots,L.
\end{equation}
The Coulomb Hamiltonian in second quantized form is given by
\begin{equation}
    \hat{H}=\sum_{ij}\langle i|h_0|j\rangle [\hat{a}_i^\dag \hat{a}_j +\hat{a}_{\bar{i}}^\dag \hat{a}_{\bar{j}}]+\frac{1}{2}\sum_{ijkl}\langle ij|V|kl\rangle [\hat{a}_i^\dag\hat{a}_j^\dag\hat{a}_l\hat{a}_k+\hat{a}_i^\dag\hat{a}_{\bar{j}}^\dag\hat{a}_{\bar{l}}\hat{a}_k+\hat{a}_{\bar{i}}^\dag\hat{a}_j^\dag\hat{a}_l\hat{a}_{\bar{k}}+\hat{a}_{\bar{i}}^\dag\hat{a}_{\bar{j}}^\dag\hat{a}_{\bar{l}}\hat{a}_{\bar{k}}],\label{app:coulomb}
\end{equation}
in which $\langle i|h_0|j\rangle$ and $\langle ij|V|kl\rangle$ are the one- and two-body matrix elements in the spatial orbital basis respectively, and the operators are expanded explicitly in their spin components.  

In second quantization, the spin operators for a pair of spin-orbitals $(i,\bar{i})$ are given by
\begin{equation}
    \hat{S}_i^+ = \hat{a}_i^\dag \hat{a}_{\bar{i}},\quad
    \hat{S}_i^-      = \hat{a}_{\bar{i}}^\dag \hat{a}_{i},\quad
    \hat{S}_i^0    = \tfrac{1}{2}(\hat{a}_{i}^\dag \hat{a}_{i}-\hat{a}^\dag_{\bar{i}} \hat{a}_{\bar{i}}),\label{app:spinoperators}
\end{equation}
with $\hat{S}_i^+$, $\hat{S}_i^-$ and $\hat{S}_i^0$ the spin raising, lowering and projection operators respectively.  The \emph{total} spin operators of a system of $L$ spatial orbitals is then given by the sum of all spin operators
\begin{equation}
    \hat{S}^+ = \sum_{i=1}^L\hat{S}_i^+,\quad
    \hat{S}^- = \sum_{i=1}^L\hat{S}_i^-,\quad
    \hat{S}^0 = \sum_{i=1}^L\hat{S}_i^0,
\end{equation}
and the magnitude of the total spin is
\begin{equation}
    \hat{S}^2=\tfrac{1}{2}(\hat{S}^+\hat{S}^-+\hat{S}^-\hat{S}^+)+(\hat{S}^0)^2.
\end{equation}
Expressed in the individual electron creation/annihilator operators, the total spin operator becomes
\begin{equation}
    \hat{S}^{2} =  \tfrac{3}{4} \sum^{L}_{i=1}(a^{\dagger}_{i} \hat{a}^{}_{i} + \hat{a}^{\dag}_{\bar{i}} \hat{a}^{}_{\bar{i}}) +  \sum^{L}_{ij=1}\hat{a}^{\dag}_{i} \hat{a}^{\dag}_{\bar{j}} \hat{a}^{}_{j} \hat{a}^{}_{\bar{i}} + \sum^{L}_{ij=1}\tfrac{1}{4} [\hat{a}^{\dag}_{i} \hat{a}^{\dag}_{j} \hat{a}^{}_{j} \hat{a}^{}_{i} - \hat{a}^{\dag}_{\bar{i}} \hat{a}^{\dag}_{j} \hat{a}^{}_{j} \hat{a}^{}_{\bar{i}} - \hat{a}^{\dag}_{i} \hat{a}^{\dag}_{\bar{j}} \hat{a}^{}_{\bar{j}} \hat{a}^{}_{i} + \hat{a}^{\dag}_{\bar{i}} \hat{a}^{\dag}_{\bar{j}} \hat{a}^{}_{\bar{j}} \hat{a}^{}_{\bar{i}}] 
\end{equation}
which has the formal structure of a one-body plus two-body operator. 

\subsection{Fock matrices}
In the spin-constrained unrestricted Hartree-Fock (c-UHF), formalism, we seek the Fock matrix of the constrained Hamiltonian 
\begin{equation}
    \hat{\mathcal{H}}=\hat{H}+\lambda (\hat{S}^2 - S(S+1)).
\end{equation}
In unrestricted Hartree-Fock (UHF) theory, only the same-spin one-body density matrices are assumed to be non-zero,
\begin{align}
    &\rho_{ij}=\langle\textrm{UHF}|\hat{a}_i\hat{a}_j|\textrm{UHF}\rangle\neq0,\quad  \rho_{\bar{i}\bar{j}}=\langle\textrm{UHF}|\hat{a}_{\bar{i}}\hat{a}_{\bar{j}}|\textrm{UHF}\rangle\neq0,\\
    &\rho_{i\bar{j}}=\langle\textrm{UHF}|\hat{a}_i\hat{a}_{\bar{j}}|\textrm{UHF}\rangle=\rho_{\bar{i}j}=\langle\textrm{UHF}|\hat{a}_{\bar{i}}\hat{a}_j|\textrm{UHF}\rangle=0,
\end{align}
with $|\textrm{UHF}\rangle$ the UHF state, leading to the non-zero components of the Coulomb Fock matrix

\begin{align}
    &F_{jl}= \langle j|h_0|l\rangle+\sum_{ik}[\langle ij|V|kl\rangle(\rho_{ik}+\rho_{\bar{i}\bar{k}})-\langle ij|V|lk\rangle\rho_{ik}],\label{app:fock:coulomb1}\\
    &F_{\bar{j}\bar{l}}=\langle j|h_0|l\rangle+\sum_{ik}[\langle ij|V|kl\rangle(\rho_{ik}+\rho_{\bar{i}\bar{k}})-\langle ij|V|lk\rangle\rho_{ik}].\label{app:fock:coulomb2}
\end{align}

The total spin operator has the structure of a one-body plus two-body Hamiltonian, which is amenable for a Hartree-Fock mean-field treatment.  Consequently the UHF Fock matrix of the constrained Hamiltonian (\ref{eq:Lag}) breaks down into the following non-zero components
\begin{align}
    &\mathcal{F}_{jl}=F_{jl}+\lambda[\tfrac{3}{4}\delta_{jl}+\tfrac{1}{2}\sum_{k}(\rho_{kk}-\rho_{\bar{k}\bar{k}})\delta_{jl}-\tfrac{1}{2}\rho_{jl}-\rho_{\bar{j}\bar{l}}],\\
    &\mathcal{F}_{\bar{j}\bar{l}}=F_{\bar{j}\bar{l}}+\lambda[\tfrac{3}{4}\delta_{jl}+\tfrac{1}{2}\sum_{k}(\rho_{\bar{k}\bar{k}}-\rho_{kk})\delta_{jl}-\tfrac{1}{2}\rho_{\bar{j}\bar{l}}-\rho_{jl}],
\end{align}
with $F_{jl}$ and $F_{\bar{j}\bar{l}}$ the Fock matrix elements (\ref{app:fock:coulomb1}-\ref{app:fock:coulomb2}) of the Coulomb Hamiltonian (\ref{app:coulomb}).  

\subsection{Spin properties of UHF}\label{section:appendix:timereversal}
There are many ways to prove spin-symmetric properties or perform spin projection \cite{lowdin1955quantum,peierls1957collective,lykos1963discussion} for two-electron systems in the literature, however we will base our argument on the strong connections with the time-reversal properties of spin.  The time-reversal operator for a collection of spin-$\frac{1}{2}$ particles is given by \cite{heyde:1994}  
\begin{equation}
    \hat{\mathcal{T}}=e^{i\pi\hat{S}_y}\hat{K} 
\end{equation}
with $\hat{S}_y$ the total spin projection in the $y$-direction and $\hat{K}$ the complex conjugation operator.  As all orbitals are real-valued, the complex conjugation operator reduces to the identity operator.  Because the spin operators commute for different spin orbitals, the time-reversal operator breaks down into a product of individual spin-orbital operators
\begin{equation}
    \hat{\mathcal{T}}=e^{i\pi\sum_{i=1}^L\hat{S}_{y,i}}\hat{K}=\prod_{i=1}^L e^{i\pi\hat{S}_{y,i}}\hat{K}.
\end{equation}
Using the definition of the individual spin operators (\ref{app:spinoperators}), each individual spin-orbital operator reduces to
\begin{equation}
    e^{i\pi\hat{S}_{y,i}}=e^{\frac{\pi}{2}(\hat{a}^\dag_{i}\hat{a}_{\bar{i}}-\hat{a}^\dag_{\bar{i}}\hat{a}_i)}=1+(\hat{a}^\dag_{i}\hat{a}_{\bar{i}}-\hat{a}^\dag_{\bar{i}}\hat{a}_i)+(\hat{a}^\dag_{i}\hat{a}_{\bar{i}}-\hat{a}^\dag_{\bar{i}}\hat{a}_i)^2,
\end{equation}
So the time-reversal operator becomes
\begin{equation}
    \hat{\mathcal{T}}= \prod_{i=1}^L[1+(\hat{a}^\dag_{i}\hat{a}_{\bar{i}}-\hat{a}^\dag_{\bar{i}}\hat{a}_i)+(\hat{a}^\dag_{i}\hat{a}_{\bar{i}}-\hat{a}^\dag_{\bar{i}}\hat{a}_i)^2]\hat{K}.
\end{equation}
For completeness, the inverse time reversal operator can be found using the same methodology
\begin{equation}
    \hat{\mathcal{T}}^{-1}=\hat{K}e^{-i\pi\hat{S}_{y,i}}=\prod_{i=1}^L[1-(\hat{a}^\dag_{i}\hat{a}_{\bar{i}}-\hat{a}^\dag_{\bar{i}}\hat{a}_i)+(\hat{a}^\dag_{i}\hat{a}_{\bar{i}}-\hat{a}^\dag_{\bar{i}}\hat{a}_i)^2].
\end{equation}
The action of the time-reversal operator on the single-particle creation and annihilation operators are given by
\begin{align}
&\hat{\mathcal{T}}\hat{a}_i^\dag \hat{\mathcal{T}}^{-1}=-\hat{a}_{\bar{i}}^\dag,\quad \hat{\mathcal{T}}\hat{a}_{\bar{i}}^\dag \hat{\mathcal{T}}^{-1}=\hat{a}_{\bar{i}}^\dag,\quad\hat{\mathcal{T}}\hat{a}_i \hat{\mathcal{T}}^{-1}=-\hat{a}_{\bar{i}},\quad \hat{\mathcal{T}}\hat{a}_{\bar{i}} \hat{\mathcal{T}}^{-1}=\hat{a}_{\bar{i}},
\end{align}
assuming the effective role of \emph{spin-flip} operator for single particles, up to a phase.  
Furthermore, it is straightforward to verify that the non-relativistic Coulomb Hamiltonian in the absence of a magnetic field (Eq.\ (\ref{app:coulomb})) is time-reversal invariant
\begin{equation}
    \hat{\mathcal{T}}\hat{H}\hat{\mathcal{T}}^{-1}=\hat{H},
\end{equation}
as is the particle number vacuum state
\begin{equation}
    \hat{\mathcal{T}}|0\rangle=|0\rangle.
\end{equation}
These are the necessary ingredient in showing that the NOCI will always restore spin symmetry for two-electron problems provided the dual state $|\overline{\textrm{c-UHF}}\rangle$ for each $|\textrm{c-UHF}\rangle$ state is included in the NOCI basis.  Indeed, it suffices to show that the $|\overline{\textrm{c-UHF}}\rangle$ and $|\textrm{c-UHF}\rangle$ are degenerate.  First, we realize that the dual $|\overline{\textrm{c-UHF}}\rangle$ state is related to the original $|\textrm{c-UHF}\rangle$ via the time-reversal operator
\begin{equation}
    \hat{\mathcal{T}}|\textrm{c-UHF}\rangle=|\overline{\textrm{c-UHF}}\rangle.
\end{equation}
Second, we see that 
\begin{equation}
    \langle\textrm{c-UHF}|\hat{H}|\textrm{c-UHF}\rangle=\langle\textrm{c-UHF}|\hat{\mathcal{T}}^{-1}\hat{\mathcal{T}}\hat{H}\hat{\mathcal{T}}^{-1}\hat{\mathcal{T}}|\textrm{c-UHF}\rangle=\langle\overline{\textrm{c-UHF}}|\hat{H}|\overline{\textrm{c-UHF}}\rangle,
\end{equation}
in which $\hat{\mathcal{T}}^{-1}=\hat{\mathcal{T}}^\dag$ has been used.  Because of this degeneracy, our NOCI Hamiltonian can always be block-diagonalized by means of a unitary transformation containing $2\times2$ Hadamard block matrices connecting the $|\textrm{c-UHF}\rangle$ and their $|\overline{\textrm{c-UHF}}\rangle$ dual states,
\begin{equation}
    \left(\begin{array}{cccc}\ddots & .  & . & . \\
    . & \frac{1}{\sqrt{2}}  & \frac{1}{\sqrt{2}}  & . \\
    . & \frac{1}{\sqrt{2}} & -\frac{1}{\sqrt{2}} & . \\
    . & . &  . & \ddots \end{array}\right)\left(\begin{array}{cccc}\ddots & .  & . & . \\
    . & \varepsilon_i  & v_i  & . \\
    . & v_i & \varepsilon_i & . \\
    . & . &  . & \ddots \end{array}\right)\left(\begin{array}{cccc}\ddots & .  & . & . \\
    . & \frac{1}{\sqrt{2}}  & \frac{1}{\sqrt{2}}  & . \\
    . & \frac{1}{\sqrt{2}} & -\frac{1}{\sqrt{2}} & . \\
    . & . &  . & \ddots \end{array}\right)=\left(\begin{array}{cccc}\ddots & .  & . & . \\
    . & \varepsilon_i+v_i  & 0  & . \\
    . & 0 & \varepsilon_i-v_i & . \\
    . & . &  . & \ddots \end{array}\right)
\end{equation}
with $\varepsilon_i=\langle\textrm{c-UHF}_i|\hat{H}|\textrm{c-UHF}_i\rangle=\langle\overline{\textrm{c-UHF}}_i|\hat{H}|\overline{\textrm{c-UHF}}_i\rangle$ and $v_i=\langle\textrm{c-UHF}_i|\hat{H}|\overline{\textrm{c-UHF}}_i\rangle$ the diagonal and off-diagonal matrix elements of the $i$th c-UHF state respectively.  For two-electron systems, these blocks can be identified as the singlet- and triplet state blocks. 

\end{document}